\documentclass[prl,reprint]{revtex4-1}
\usepackage{graphicx,epstopdf}
\usepackage{psfrag}
\usepackage{epsfig}
\usepackage{color}
\usepackage{multirow}
\usepackage{amssymb}
\graphicspath{{./}{pics/}}
\usepackage{enumerate}

\draft % marks overfull lines with a black rule on the right

\begin{document}




\preprint{}

\title{Experimental investigation of an oil droplet colliding with an oil-water interface} %Title of paper








\author{U. Miessner}
\email[]{U.Miessner@tudelft.nl}
% \homepage[]{http://ahd.tudelft.nl/~gosse/}
\author{E. Coyajee}
\author{R. Delfos}
\author{R. Lindken}
\author{J. Westerweel}

%\thanks{}


\affiliation{Laboratory for Aero and Hydrodynamics, Leeghwaterstraat 21
2628 CA Delft, The Netherlands}





\date{\today}

% word count first version: 7672 words, 48385 characters. 
% first vrevised version 1p1 has 7860 words.

\begin{abstract}

The impact of a buoyancy driven oil droplet with an oil-water interface
is investigated using time-resolved Particle Image Velocimetry (PIV)
along with a phase discrimination by means of high-speed Laser Induced
Fluorescence (LIF). In this paper we focus on the investigation of strategies
to optimize the performance of high-speed PIV algorithms. Furthermore this
data will be used for validation of numerical simulations of two phase flows.
In order to simultaneously measure the flow velocities inside and around
the oil droplet by means of PIV the refractive indices of both phases need to
be carefully matched. The aqueous phase consists of a mixture of corn syrup
and water, which defines the viscosity as well as the refractive index. The
disperse phase consists of a mixture of two kinds of mineral oils. The latter
are mixed to match the refractive index of the continuous phase. Both phases
are seeded with tracer particles required for PIV. A fluorescent dye is added
to the dispersed phase to allow discrimination of the PIV signals originating
from both phases by means of LIF.
The LIF and PIV signals are captured by two aligned, synchronized highspeed
cameras, each used for one of the measurement techniques. In order to
study the impact of an oil droplet into an oil-water interface, it is necessary
to accurately measure a velocity range that is spread over three orders of
magnitude. The impact of the droplet is measured in a time series of PIV
recordings with a high temporal resolution. This allows to optimize the time
interval between two correlated frames within the time series to achieve a
high signal-to-noise ratio, while still being able to measure a large velocity
dynamic range. An approach to estimate an optimal time separation is presented
in this study. The results are compared to the experimental data of Mohamed-Kassim and Longmire \cite{mohamed2003drop} and the effectiveness of the introduced optimization on the data quality is discussed. Furthermore the results of this
experiment are compared to a numerical simulation developed by Coyajee et al. \cite{coyajee2006direct}.

\end{abstract}

% 47.55.df
% Breakup and coalescence
% 47.55.dr
% Interactions with surfaces
% 68.05.Cf
% Liquid-liquid interface structure: measurements and simulations
% 68.03.Cd
% Surface tension and related phenomena

\maketitle %\maketitle must follow title, authors, abstract and \pacs



%======================================================================================
\section{Introduction}
\label{sec:Introduction}
%======================================================================================

A large variety of physical and chemical processes involve two immiscible
liquids, e.g. extraction, oil production and transport, emulsification and separation.
Understanding of the motion of individual droplets and their interactions
is crucial for economically and ecologically optimized design. While
the behavior of single droplets is known to a large extent, the interaction of
two dispersed droplets (e.g. deformation, break-up, collision and coalescence)
is less understood.
In the past, several studies have investigated the interaction of droplets
and surfaces in order to understand the complex mechanisms involved. A
number of studies dealt with the examination of droplets falling through a gas
before impacting on a liquid surface. For an overview about PIV experiments
in dispersed two-phase flows the reader is referred to \cite{mohamed2003drop}, \cite{delfoskaijenjet2}, \cite{delfoskaijenjet}.
Because of the low viscosity of the surrounding gas the film between the
impacting droplet and the surface drains quickly in this case: the droplet
immediately coalesces. To study impact without instant merge of the two
liquids, a higher viscosity of the surrounding fluid is required to establish a
thin film between the surface and the droplet. Coalescence will eventually
occur once the film is drained, but it is no longer directly coupled to the
impact event.
Mohamed-Kassim and Longmire \cite{mohamed2003drop} published an experimental study
using this approach, focusing their investigation on the hydrodynamics of
the liquids involved in the impact. The viscosity ratio chosen is much closer
to unity the ratio realized in this study. The data obtained in this study will
provide a more demanding benchmark for numerical simulations. High-speed
PIV (\cite{westerweelrecent}, \cite{narrow2000simple}, \cite{hainkahler}) with a frequency of 500 Hz was applied to visualize and
quantify the fluid flow. However, it is possible to further increase the accuracy
of such measurements with a higher frame rate and a method which allows to
adjust the time interval between two correlated particle images to the actual
change of flow conditions.
In this work the highly instationary flow conditions during a droplet impact
are chosen to provide a laminar test case with steep spatial and temporal
velocity gradients. Performing measurements in such a flow at very high sample
frequencies does not automatically provide better results if consecutive
particle images are correlated: the measured displacement, which scales with
the delay time between two images, can become so small that they will be
of the same order of magnitude as the measurement accuracy of the system.

While the correlation of the particle images \footnote[1]{Expressed as the signal-to-noise ratio (SNR), usually obtained from the correlation peak height.} may be high (leading to a low
percentage of ‘spurious’ vectors), the result will suffer from a lack in dynamic
range and will thus be noisy. Increasing the time separation between two images
decreases the signal-to-noise ratio (SNR) but simultaneously increases
the velocity dynamic range (VDR). Depending on the flow conditions, each
image pair requires an optimal time separation in order to increase the overall
accuracy of the investigation.
The aim of the present study is to identify parameters which can be
utilized to adjust the separation time individually for each correlation. Additionally,
it is our objective to use the data to validate results from numerical
simulations. Thus a higher viscosity ratio between continuous and dispersed
phase is chosen to provides a demanding test case for numerical simulations.
The LIF data is used to detect and track the interfaces between the phases.
The experimental results are then compared to numerical results from a simulation
performed with the recently developed Mass Conserving Level-Set
(MCLS) method \cite{vdpijlthesis}, \cite{van2005mass}. The detailed validation will be topic of a different
paper.

%======================================================================================
\section{Experimental}
%======================================================================================

\subsection{Facility}
\begin{figure}
\includegraphics[scale=0.4]{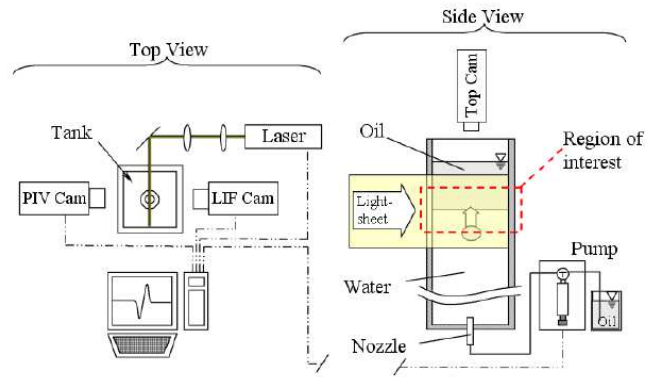}
\caption{Schematic representation of the experimental set up showing the top view
(left) and the side view (right) of the tank and measurement equipment.}
\label{pic1}
\end{figure}

A glass tank with a square section of 50 mm width and 600 mm height is
used for the investigation of the droplet impact (see Fig. \ref{pic1}). Between nozzle
and interface a distance of 100 mm is maintained. The liquid layer on top of
the bulk phase has a thickness of 50 mm.
The liquids used are a glucose-water mixture (C6H12O6/H2O) for the
continuous phase and an oil-oil mixture (Shell Macron EDM 110 / Shell
Garia GX 32) for the dispersed phase. Values for the liquid properties are
given in table 1. The refractive index of the mixtures is matched to minimize
optical distortion with an accuracy of the actual value of 0.03\%. Droplet
and top layer consist of the same liquid. A more detailed description of the
matching procedure is given in section 2.2. A droplet with an equivalent diameter of D = 11 mm is generated by
injecting the oil mixture into the bulk phase through a cylindrical nozzle
of 60 mm length and 5 mm diameter. To ensure that this process is reproducible,
the injection is driven by a computer-controlled syringe pump.
Tracer particles (‘Sphericell 110-P-8’) are added to both phases. The tracer
particles have a nominal diameter of 10 μm to visualize and measure the
flow conditions with PIV. Additionally, a fluorescent dye (‘Hostasol Yellow
3G’) is dissolved into the oil mixture in order to discriminate between the
two refractive-index-matched liquids.
A thin planar light sheet (thickness ∼ 1 mm) is formed using a pulsed
Nd:YLF laser (527 nm, New Wave ’Pegasus’) in combination with a cylindrical
and a spherical lens (focal lengths: Fc = -80 mm, Fs = 350 mm).

The laser light sheet illuminates a measurement volume in the middle of the
tank of approximately 25 $\times$ 50 mm, centered around the impacting droplet.
Two high-speed cameras (Photron: Ultima APX-RS, 1024 $\times$ 1024 pixels) are
mounted orthogonally to the measurement plane (one on each side, see Fig.
\ref{pic1}). They are triggered synchronously with the laser pulses; the laser and
cameras are operated at 3000 Hz. A third camera (Sensicam QE) is mounted
above the experimental section (’Top Cam’ in Fig. \ref{pic1}) to validate the droplet
position relative to the light sheet \cite{lindken2000velocity}.
Alignment of the high-speed cameras is done by cross-correlation particle
images of camera 1 with camera 2 \cite{willert1997stereoscopic} (both recorded at the same instance).
An optical filter (Schott ref.$\#$OG570, $\lambda_{cut−off} = 570 \pm 6$ nm) is mounted in
front of camera 2.The orange filter absorbs the light at the original laser wave
length (532 nm) and transmits the longer wavelength of the fluorescence.
Camera 2 records the scattered light from the tracer particles through a gray
filter (B+W Graufilter, neutral density 2). The gray filter is used to reduced
the light intensity towards camera 2 in such a way, that the LIF recordings
can be illuminated with the same laser pulse of about 100 mJ energy. The
syringe pump is coupled to the timing unit of the PIV-system. The data
acquisition can be triggered by the droplet injection controller.

The experimental conditions mentioned above are comparable to those
reported in \cite{mohamed2003drop}. Important differences between both experiments are the
restricted domain size and the motion of the impacting droplet (sinking
droplets vs. rising droplets in the current study). The reason for the domain
restriction is the correspondence to the numerical simulation which is
computed in an identical finite domain. Due to symmetry reasons, the opposite
droplet motion in the set-up does not have significant effects when
the results are compared to the results presented by \cite{mohamed2003drop}. A list of physical
properties of both investigations is given in table 1. Here $U_i$ is the impact velocity,
$t_g = \sqrt{D/(\Delta \rho \cdot g/\rho_{mean})}$ denotes the gravity timescale and $t_i = D/U_i$
is the impact time scale. The Weber number $We = \rho_d U_{i}^{2} /\sigma$ , where $\rho_d$ is
the density of the droplet, D the droplet diameter and $\sigma$ the surface tension,
and Reynolds number $Re = \rho_s U_{term}D/\mu_s$ use the index ’s’ for ambient
fluid and ’d’ for droplet fluid. The density ratio in Tab. 1 is given as a reciprocal
because of the inversed droplet motion. Due to the choice of liquid
mixtures the viscosity ratio is much lower than the experiment of \cite{mohamed2003drop}. Also a
higher Weber number occurs because of the lower surface tension coefficient
$\sigma = 0.02N/m$.

\begin{table}[h!]
\caption{Experimental conditions.}
\begin{center}
\begin{tabular}{ l  l  l  l  l }
\hline
\multicolumn{1}{c}{} & \multicolumn{3}{c}{Mohamed-Kassim \& Longmire} & experiment/DNS \cite{} \\ 
\hline
$D$ & cm & 1.03 & 1.03 & 1.1 \\
$U_i$ & cm/s & 13.2 & 9.8 & 25 \\
$t_i$ & ms & 78 & 105 & 44 \\
$t_g$ & ms & 78 & 80 & 49 \\
$\rho_d/\rho_s$ & - & 1.189 & 1.178 & $(1.625)^{-1}$ \\
$\mu_d/\mu_s$ & - & 0.33 & 0.14 & 0.03 \\
$Re$ & - & 68 & 20 & 36 \\
$We$ & - & 7.0 & 3.8 & 28 \\
$Fr$ & - & 1.0 & 0.6 & 0.6 \\
\hline \end{tabular}
\label{table:timetocoal}
\end{center}
\end{table}

\subsection{Refractive index matching}
\label{sec:PIV experimental method}

\begin{figure}
\includegraphics[scale=0.4]{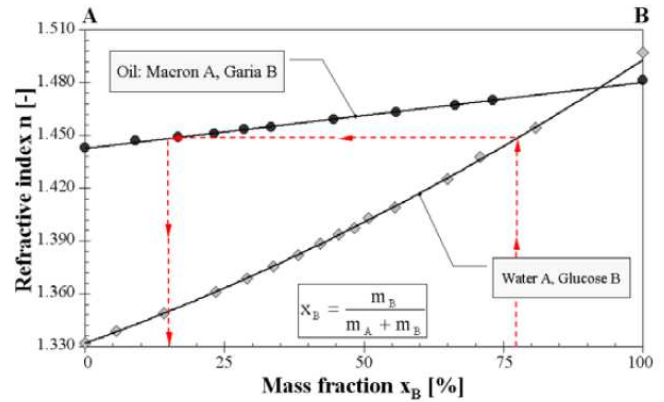}
\caption{Refractive index vs. mass fraction xB at a fixed temperature of T = 298
K ($25^{\circ}{\rm C}$). The achievable matching range of refractive indices is restricted due to
range of the oil phase. Therefore only aqueous mass fractions between 74\% and
93\% may be used for index matching.}
\label{pic2}
\end{figure}

PIV measurements are performed inside and outside the impacting droplet.
A mismatch of refractive indices would lead to optical distortion due to a
lens effect caused by the curvature of the droplet interface. To avoid optical
distortions the refractive indices of the dispersed and the continuous phase
have to be matched \cite{zachos1996piv}.
Recently a review of index matching methods of liquid mixtures and
solid materials has been published \cite{miller2006matching}. When the refractive index of a liquid
mixture to a solid material is matched one degree of freedom is necessary.
However, in this experiment both the refractive index and the viscosity ratio
need to be adjusted. The two degrees of freedom are produced by matching
an oil/oil mixture to a water/glucose mixture.
The refractive indices of the mixtures are dependent on the mixing ratio of
the components as well as on the temperature. Thus, the influence of mixing
ratio and temperature is investigated in two separate experiments. First, at
a fixed temperature we determine the range in which the refractive indices
of both mixtures can be adjusted to one another (Fig. \ref{pic2}). In the second
experiment the analysis of the change of viscosity and refractive index with
temperature is then used to interpolate the temperature behavior of the
refractive index of different mass fractions (see Fig. \ref{pic4}).
To reveal the range of matchable indices at a fixed temperature (Fig. \ref{pic2})
the mass fraction $x_B = m_B/(m_A + m_B)$ of the aqueous and the oil mixture
is varied and the resulting refractive index is measured. The refractive index
range is determined by the oil mixture due to its smaller range (1.4431 -
1.4813). Thus the range of mass fractions of aqueous mixtures is restricted
to values between $x_{B,aqueous}(1.4431) = 0.74$ and $x_{B,aqueous}(1.4813) = 0.93.$

\begin{figure}
\includegraphics[scale=0.4]{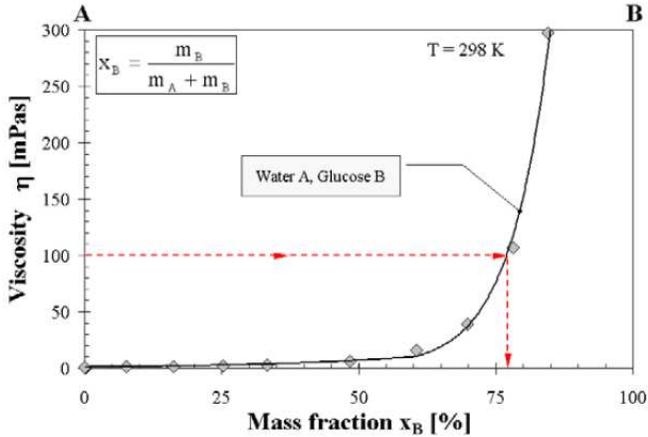}
\caption{Dynamic viscosity $\eta$ vs. mass fraction $x_B$ of glucose B mixed into the water
phase A. To avoid rapid droplet breakup, the viscosity of the aqueous mixture is
chosen to be 100 mPas. The resulting mass fraction is therefore 77\%.}
\label{pic3}
\end{figure}

\begin{figure}
\includegraphics[scale=0.4]{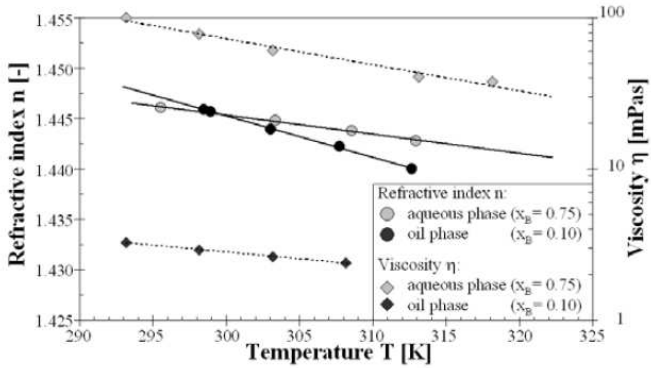}
\caption{Refractive index and dynamic viscosity vs. temperature at a fixed mass
fraction $x_B = 10\%$ for the oil phase and $75\%$ for the aqueous phase. The relative
change of the quantities shown are considered to be independent of the mass fraction
for small deviations from the target values.}
\label{pic4}
\end{figure}

For the validation of the above mentioned simulation \cite{coyajee2006direct} a droplet impact
without immediate breakup is needed. To produce this kind of impact, rapid
drainage of the film between the droplet and the upper surface has to be suppressed.
A viscous continuous phase will ensure slow thinning of the film between
droplet and top layer. In Fig. \ref{pic3} the viscosity of the water-glucose mixture
is presented as a function of the mass fraction $x_B$. Referring to the range
of matchable refractive indices, possible viscosities of the aqueous phase range
between $\eta_{aqueous}(0.74) = 67 mPas$ and $\eta_{aqueous}(0.93) = 934 mPas$. The viscosity
of the continuous phase is chosen to be $\eta_{aqueous}(0.77) = 100$ mPas.
The corresponding refractive index is according to Fig. \ref{pic2} $n(x_{B,aqueous} = 0.77)
= 1.4482$. Thus, the oil mixture has to have a mass fraction $x_{B,oil} = 0.15$
to match the refractive index. Determining the viscosity of the oil mixture
$\eta_{oil}(0.15) \approx 3 mPas$ leads to a viscosity ratio of $\eta_{aqueous}/\eta_{oil} \approx 33$.
The refractive index and the viscosity are very sensitive to variations in
temperature. This leads to the necessity to predict and compensate for these
property changes. Therefore the temperature dependencies of both the viscosity
and the refractive index are measured for different mass fractions $x_B$.
However, the temperature correction can be based on an interpolation from
a single mass fraction for each mixture, because small changes in the mass
fraction do result in an offset of the functional relation, but not in a change
of its shape. The data is presented in Fig. \ref{pic4}, showing the graphs for a mass
fraction $x_B = 0.10$ of the oil mixture and 0.75 for the aqueous phase. The
refractive index is linearly dependent on the temperature while the viscosity
depends exponentially. Based on the dependencies provided above, the
viscosity of the continuous phase at a desired temperature is chosen and

the required mass fraction $x_B$ of the refractive index matched mixtures are
calculated.
For other refractive index matching strategies the interested reader is
referred to \cite{budwig1994refractive}, \cite{narrow2000kut}, \cite{nguyen2004method}, \cite{cui1997refractive}, \cite{zachos1996piv}.

\subsection{Image processing}

The droplet impact is simultaneously observed with two high-speed cameras.
One camera records the LIF signal and the second one records the PIV
signal. Therefore the processing of the captured images from the experiments
consists of two parts: First, extraction of droplet location and shape (from
the LIF data) and second, measurement of the fluid velocity field (from the
PIV data). The PIV data is used to study the possibility of optimizing the
time separation between two correlated images.

\subsubsection{LIF processing}

\begin{figure}
\includegraphics[scale=0.3]{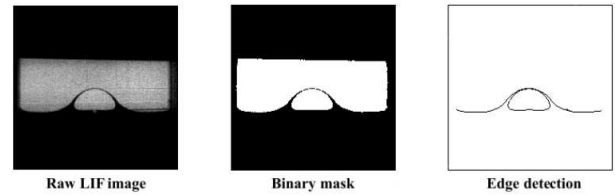}
\caption{Processing of the LIF images: A threshold filter is applied to the raw LIF
image. The resulting binary mask will provide the position of the interfaces using
an edge detection method (Sobel).}
\label{pic5}
\end{figure}

The evaluation of the LIF images is performed in Matlab 7.0. First, each
image is converted to a binary bitmap using a threshold filter (see Fig. \ref{pic5}).
The threshold value is a fixed value for the entire image series. From the
resulting binary images, the following parameters are extracted:
\begin{itemize}
 \item position of the interface
 \item the droplet top position
\item the center of area
\item the bottom of the droplet
\item horizontal and vertical diameter
\end{itemize}
Furthermore, using an edge detection method (Sobel), the shape of both
the droplet and the deforming interface is captured. This information will
also be used to identify the different phases in the PIV results.

\subsubsection{PIV processing}

The raw PIV images are pretreated for the analysis in chapter 3. Due to
extinction the intensity of the laser light sheet is not uniform over the measurement
area. These intensity gradients of the light sheet can be removed
from the raw images by applying a non-linear filter, which subtracts the
sliding minimum from the PIV image. The spatial sliding minimum filter computes the minimum intensity in a neighborhood of pixels. This is done
for efficiency using a sliding minimum filter operation. The spatial sliding
minimum is then subtracted to remove background noise. The filter length
is chosen much larger than the particle image diameter, in order to maintain
the particle information \cite{lindken2006stereoscopic}.
The PIV calculation of vector fields is performed with in-house software
by using cross-correlation with a decreasing interrogation window size starting
with 32×32 pixels to estimate the pre-shift for two subsequent correlations
using 16×16 pixel interrogation windows with 50\% overlap. The obtained
vector field is subsequently post-processed using a universal outlier detection
algorithm \cite{westerweel2005universal} to label spurious vectors.

\section{Evaluation}

During a short period of time of approximately 2 seconds, the rising droplet
approaches the interface, impacts into the oil top layer and oscillates around
its equilibrium state. Coalescence of the droplet with the top-layer is prevented
by a thin film of the continuous phase. Only after several minutes,
the thin film is drained and the droplet merges with the top layer. During the
impact the velocity magnitude changes in time and space (i.e. acceleration/
deceleration due to the oscillation and velocity gradients due to strong local
fluid motion). A fixed time interval between the correlated particle images
throughout the impact sequence will lead to inaccuracies, either in the parts
of the measurement sequence with low or with high velocities. This error
depends on the chosen length of the time interval.
In order to increase the accuracy of the high-speed PIV temporal and
spatial variations of the velocity within the sequence have to be taken into
account. Since PIV is based on cross-correlation, a measure for valid correlations
is the signal-to-noise ratio (SNR), which is given by the ratio between
the highest correlation peak to the second highest. An excellent SNR can be
achieved using the smallest time interval possible. However the measured particle
image displacements will be in the range of the measurement accuracy
of the PIV system. Applying a long time interval results in sufficient SNR for
small particle image displacements, but causes decorrelation in faster parts
of the vector field. Thus an optimum has to be found to combine a high
velocity dynamic range (VDR) with a high SNR.

\subsection{Analysis of the vector fields}

\begin{figure}
\includegraphics[scale=0.4]{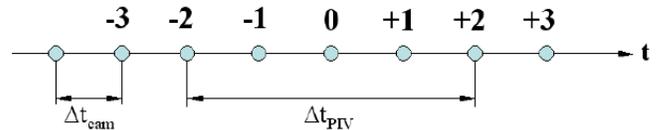}
\caption{Adjustment of time interval $\Delta t_{PIV}$ : In this example three images are
skipped between the correlated images. This results in an effective separation time
$\Delta t_{PIV} = 4 \Delta t_{cam}$ between frames $n=-2$ and $n=+2$ for the time $t0$.}
\label{pic6}
\end{figure}

\begin{figure}
\includegraphics[scale=0.4]{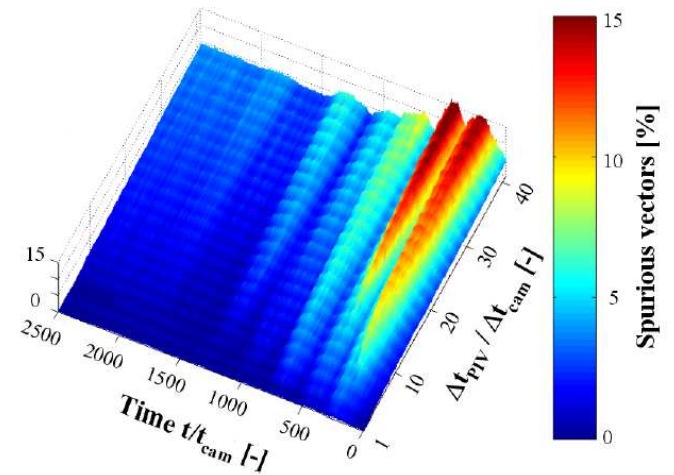}
\caption{Survey of percentage of spurious vectors over time $t/t_{cam}$ and time interval
$t_{PIV} /t_{cam}$.}
\label{pic7}
\end{figure}

In order to search for a quantity that indicates the quality of a PIV evaluation
the sequence of a single droplet impact event is evaluated several times, each
time increasing the time interval $\Delta t_{PIV} = k \cdot \Delta t_{cam}$, with $\Delta t_{cam}$ being
the time interval between consecutive pictures. For $\Delta t_{PIV} /\Delta t_{cam} = 4$ an
example is given in Fig. \ref{pic6} for a time $t_0$ during a sequence.
As a first indication of the quality of a PIV evaluation the percentage of
outliers inside each vector field of the above described data sets is calculated.
The number of outliers can be regarded as a measure of the decorrelation of
the PIV signal. According to Keane and Adrian \cite{keane1990optimization} an amount of spurious
vectors of less than 5\% in a PIV measurement is considered to be acceptable.
In Fig. \ref{pic7} the development of the outliers in the computed vector fields
is shown as a function of time and length of the time interval between correlated
particle images. Each axis is normalized with the recording interval
$\Delta t_{cam}$. The oscillation of the outlier amplitude corresponds to the oscillating
movement of the droplet during the impact. Increasing time interval $\Delta t_{PIV}$
results in a magnification of the amplitude of the oscillation (i.e. an increase
of spurious vectors).

A comparison of the occurrence of outliers for three different time intervals
(consecutive, $3\Delta t_{cam}$, $6\Delta t_{cam}$) reveals important similarities. All maxima
occur at the same time instant. The first maximum is located at $t/t_{cam} = 212$
and will be used in the following as a reference for comparison.

\begin{figure}
\includegraphics[scale=0.4]{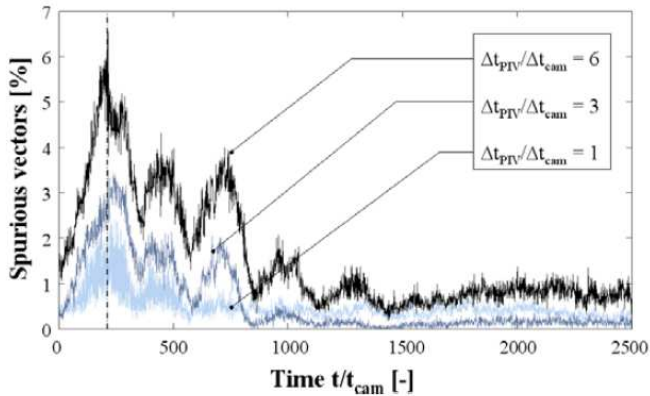}
\caption{Percentage of Spurious vectors of the time series using time interval with
$t_{PIV} /t_{cam} = 1,3$ and $6$ respectively.}
\label{pic8}
\end{figure}

\subsection{Determination of an indicator}

\begin{figure}
\includegraphics[scale=0.4]{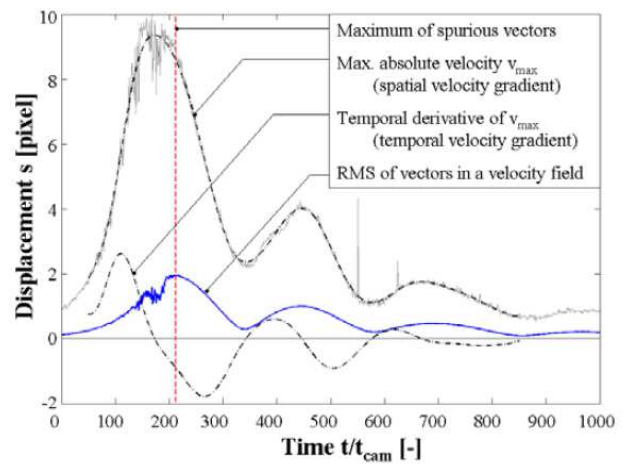}
\caption{The maximum of absolute particle image displacement as an indicator
for spatial velocity gradients is compared to the change of displacement as an
qualitative indicator for the temporal velocity gradients. The maximum of the
RMS of the vector fields coincides with the occurrence of the maximal number of
spurious vectors and will thus be used in the following for the optimization.}
\label{pic9}
\end{figure}

In order to obtain a more relevant link to the occurrence of spurious vectors
the origin of the outliers with respect to the evolving flow needs to be detected.
Possible reasons for a failure of the cross correlation are spatial and
temporal velocity gradients.
Results of a more detailed investigation are given in Fig. \ref{pic9}. The maximum
absolute displacements of the vector fields are used as an indicator for
spatial velocity gradients. The temporal derivative of this quantity serves as
a qualitative indicator for temporal velocity gradients.
With respect to the first maximum of the outlier development $t/t_{cam} =
212$, it can be shown that temporal velocity gradients (i.e temporal derivative
of the maximum particle image displacement) are not the cause of the
occurring outliers. In fact the spatial gradients (i.e. maximum particle image
displacement) are considered to be the cause of the spurious data.
A good indicator for spatial velocity gradients is given by the standard
deviation (RMS) of all displacements within a vector field. This quantity
increases with increasing differences in the spatial distribution of velocities.
In Fig. \ref{pic9} it can be seen that the maximum of the RMS value coincides with the
maximum of spurious vectors. Thus, this quantity will be used as a criterion
to optimize the time interval $\Delta t_{PIV}$ between two correlated images.

\subsection{Optimization}

\begin{figure}
\includegraphics[scale=0.4]{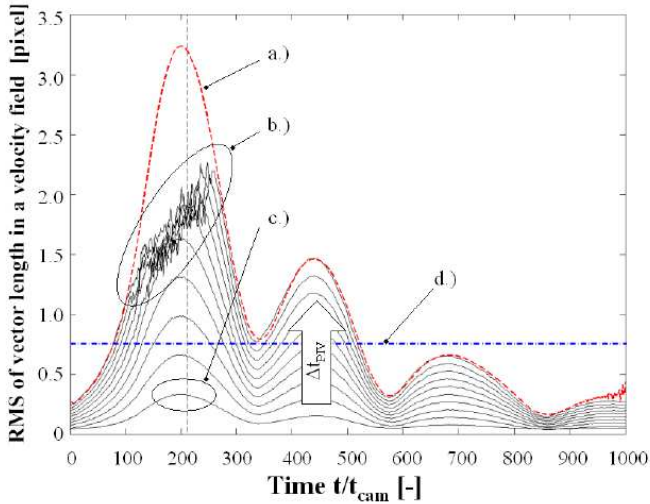}
\caption{Oscillation of the RMS of the particle image displacement (varied time
interval $t_{PIV} /t_{cam}$ for each graph 1 up to 11), influence of spurious vectors and
threshold chosen.}
\label{pic10}
\end{figure}

After identifying the RMS value of the vector fields as an indicator for the
occurrence of outliers, a connection is needed between the change of RMS to
the time interval between correlated particle images. Thus, an investigation
of the RMS of the mean particle image displacement over time and increasing
time interval $\Delta t_{PIV}$ is shown in Fig. \ref{pic10}.

\begin{figure}
\includegraphics[scale=0.4]{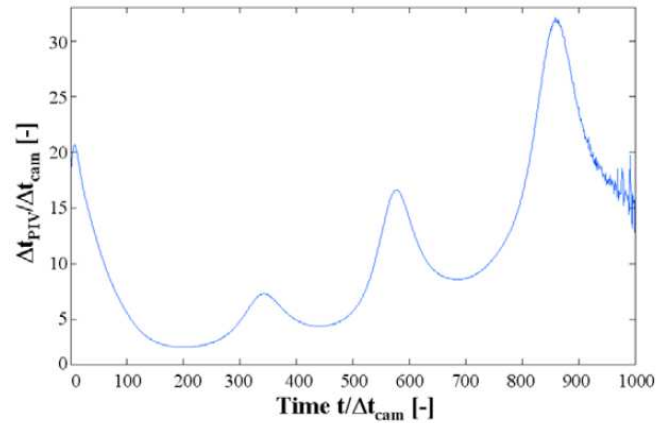}
\caption{Time interval $\Delta t_{PIV} /\Delta t_{cam}$ adjusted to the different parts of the impact
sequence.}
\label{pic11}
\end{figure}

Fig. \ref{pic10} reveals that the value of the RMS scales (obviously) linearly with
the time interval. Normalizing the RMS developments with the time interval
shows that the evolution of the standard deviation throughout the times
series is self-similar. Thus, fitting a polynomial with a least-square fit to the
normalized RMS of the displacements enables computation of the change
of the standard deviation as a continuous 2-D function with respect to the
impact time frame $t$ and time interval $\Delta t_{PIV}$ . To indicate the resemblance
of the fit, Fig. \ref{pic10} a.) shows the part of the 2-D function which represents
the time interval $\Delta t_{PIV} = 11 \cdot \Delta t_{cam}$. The deviation of the fit to the data
is caused by decorrelation. Large displacements result in a loss of pairs in
time. If that occurs, closer particle image pairs are correlated and lead thus
to smaller displacements (Fig. \ref{pic10} b.)).
In order to avoid decorrelation the optimized time interval needs to be
adjusted such that the resulting RMS of displacements do not indicate the occurrence
of spurious vectors. However, the time interval has to be sufficiently
long to keep the measured displacements outside the measurement accuracy
of the PIV system. Fig. \ref{pic10} c.) shows the RMS development of the data set
for $\Delta t_{PIV} /\Delta t_{cam}=1$, i.e. consecutive image frames and therefore serves as
lower boundary for the optimization. The upper limit is determined by the
occurrence of spurious vectors. Setting the target RMS of the displacements
to a constant value (Fig. \ref{pic10} d.))results in a contour line of the 2-D fit and
provides the necessary functional dependence between the impact time frame
t and the time interval $\Delta t_{PIV}$.
In Fig. \ref{pic11} the dependence of the time interval of the correlation $\Delta t_{PIV}$
and the impact time $t/t_{cam}$ is plotted. Sections in the sequence where the

fluid is moving fast have a short time interval (see $t/\Delta t_{cam} = 212$, i.e. small
$\Delta t_{PIV} /\Delta t_{cam}$). When the liquid moves slowly the displacements will be calculated
with a large $\Delta t_{PIV}$ ($t/\Delta t_{cam} = 850$, i.e. large $\Delta t_{PIV} /\Delta t_{cam}$ ).
An optimized time series of velocity data is generated by choosing the
evaluation with the optimal time interval according to Fig. \ref{pic11} for each individual
time step during the sequence. For that purpose the continuous
contour line of optimal time intervals in Fig. \ref{pic11} needs to be transferred in
discrete integer values. All measured displacements are normalized with their
original time interval $\Delta t_{PIV}$ .

\section{Results and discussion}

This chapter is divided into two parts. First, the result of the optimization is
presented and discussed. Second, a brief comparison with the corresponding
numerical simulation \cite{coyajee2006direct} is given.

\subsection{Optimization}

\begin{figure}
\includegraphics[scale=0.4]{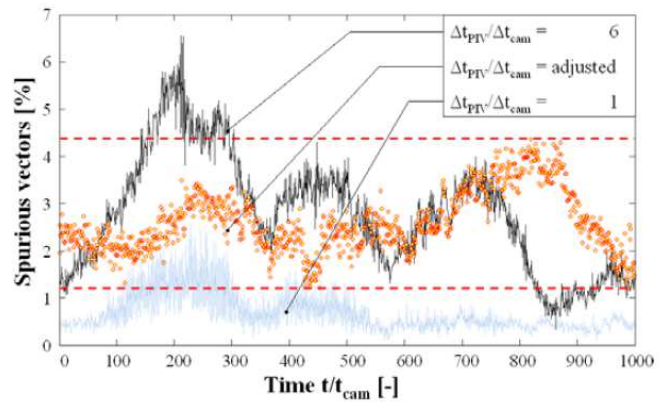}
\caption{Percentage of spurious vectors with adjusted time interval $\Delta t_{PIV} /\Delta t_{cam}$
compared to the results of a the series evaluated with $\Delta t_{PIV} /\Delta t_{cam} [1, 6]$.}
\label{pic12}
\end{figure}

Due to the discrete nature of the variability of the time interval the rounding
procedure used does not exactly produce a constant fraction of spurious vectors
throughout the time series. In Fig. \ref{pic12} the result for the spurious vectors
of the optimized series is compared to the spurious vector occurrence for a series
with fixed time separations of $\Delta t_{PIV} /\Delta t_{cam} = 1$ and $6$ respectively. The
optimized time series shows that the amount of spurious vectors is reduced 
below the limit of 5\%. The time varying behavior on the flow conditions
cannot be recognized any longer and the percentage of spurious vectors is
limited to an almost constant value between 1.2 and 4.4\% (dashed lines).

\begin{figure}
\includegraphics[scale=0.3]{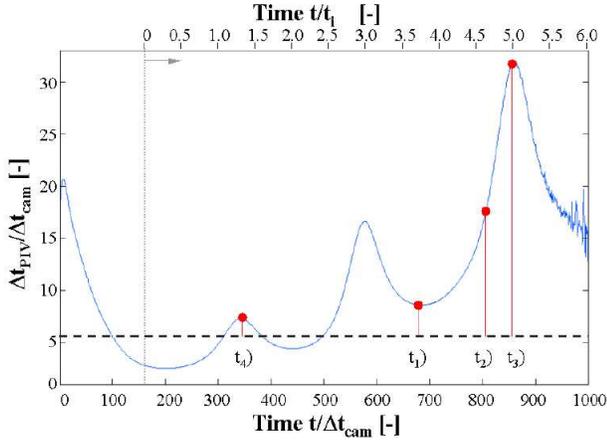}
\caption{The optimized time interval is plotted in comparison to a constant time
interval. t1), t2) and t3
) respectively show the time instants at which vorticity
plots of both approaches are compared. On t4) a minimum of the aspect ratio
occurs. This time instant is chosen for a comparison with the vorticity features
discovered by \cite{mohamed2003drop}. For convenience the time scale presented by \cite{mohamed2003drop} is introduced,
where ti = D/Ui represents an impact velocity time scale based on the droplet
diameter D and the impact velocity Ui of the droplet.}
\label{pic13}
\end{figure}

Since the vorticity is obtained by differentiation it is sensitive to spurious
vectors. Thus, to show differences between a constant PIV evaluation
time interval and the above presented optimized approach, vorticity plots are
compared. In Fig. \ref{pic13} the change in length of the time interval $\Delta t_{PIV} /\Delta t_{cam}$
is plotted vs. the impact time scale $t/\Delta t_{cam}$. In order to compare the time
scale used by \cite{mohamed2003drop} with this experiment a second time scale $t/t_i$ is introduced.

Here $t_i = D/U_i$ represents an impact velocity time scale based on the droplet
diameter $D$ and the impact velocity $U_i$ of the droplet.
The dashed line represents a PIV evaluation using a constant time interval
$\Delta t_{PIV} = 6\Delta t_{cam}$. The length of constant time interval is chosen to be the
same as in \cite{mohamed2003drop}. To show differences in the resulting vorticity fields three time
instances $t_1)$, $t_2)$ and $t_3)$ are compared. The positions represent part of an
oscillation between an upper and a lower turning point of the droplet. At time
$t/t_i = 3.1$ the droplet starts to move from the lower turning point. When the
maximum velocity is reached, the time interval is minimal (time instant $t_1)$
in Fig. \ref{pic13}). With time $t_2$ at $t/t_i = 4.9$ in Fig. \ref{pic13} an intermediate position
is chosen between the maximal fluid movement and the upper turning point
$t_3$ at $t/t_i = 5.3$. The aspect ratio of the droplet has a first minimum at
$t/t_i = 1.9$ ($t_4$
in Fig. \ref{pic13}). This time instant will be used to compare the
physical features of this flow with the results published by \cite{mohamed2003drop}.

\begin{figure}
\includegraphics[scale=0.3]{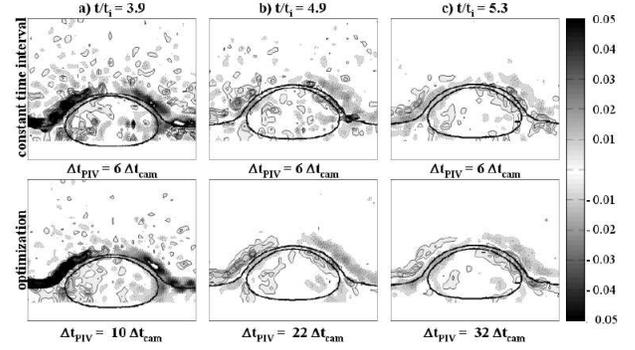}
\caption{Comparison of PIV evaluation with a constant time interval
$\Delta t_{PIV} /\Delta t_{cam} = 6$ (top row) with the optimized approach (bottom row). The time
instants shown agree with the time instances $t_1)$, $t_2)$ and $t_3)$
in Fig. \ref{pic13}. Positive
vorticity values are represented by solid lines and negative by dashed lines.i Due
to the optimization the vorticity plots are less noisy and more symmetric}
\label{pic14}
\end{figure}

\begin{figure}
\includegraphics[scale=0.3]{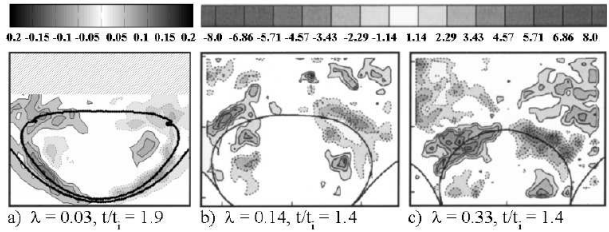}
\caption{The vorticity flow features published by Mohamed-Kassim and Longmire
\cite{mohamed2003drop} with two different viscosity ratios $\lambda = 0.14$ and 0.33 at the impact time
$t/t_i = 1.4$ compared with the according result of the present study ($\lambda = 0.03$,
$t/t_i = 1.9$). Positive vorticity values are represented by solid lines and negative by
dashed lines.}
\label{pic15}
\end{figure}

From the data sets in Fig. \ref{pic14} ($t_1$, $t_2$, $t_3$) the vorticity plots in Fig. \ref{pic15} are
computed. For all three time instants the evaluation using the constant time
interval $\Delta t_{PIV} /\Delta t_{cam} = 6$ is shown in the top row and the optimized time
interval is printed below. The deviation of fixed and variable time interval increases with time $t/t_i$ from a) to c) due to the reduction of velocity from a
local maximum a) to the upper dead point c) of the oscillation cycle.
In Fig. \ref{pic14} a) the time interval differs by a factor of 1.6 due to the higher
velocity of the flow. The vorticity plot of the adjusted time interval contains
less noise and the flow features are represented more clearly than in the constant
approach. The symmetry is more explicit in the optimized evaluation.
Differences in terms of noise are most obvious in the intermediate example
in Fig. \ref{pic14} b) when the flow slows down towards the upper dead point. The
optimization has achieved reduction of noise while the features of the flow
are captured as well as in the constant approach. The symmetry plane is
more expressed in the optimized case. Due to the minimal movement in the
upper dead point (Fig. \ref{pic14} c)) the magnitude of vorticity in both evaluations
is smaller than in the first examples. However the noisier appearance of the
vorticity due to the constant time interval is still visible in comparison with
the optimization. The symmetry of the flow is more expressed in the adjusted
approach.

The drop internal circulation at a characteristic instant during the impact
of the droplet is shown in Fig. \ref{pic15} a) to c). Based on these plots, the physical
features of the experiments of Mohamed-Kassim and Longmire \cite{mohamed2003drop} are
compared to this study. In Fig. \ref{pic15} b) and c) two vorticity plots of \cite{mohamed2003drop} with
a viscosity ratio of $\lambda = 0.14$ and $\lambda = 0.33$ respectively are shown. The time
instant chosen is $t/t_i = 1.4$ due to the occurrence of the first minimum of
the droplet aspect ratio. In the present study (Fig. \ref{pic15} a)) the first minimum
of the aspect ratio occurs at a later time instant $t/t_i = 1.9$ and the viscosity
ratio $\lambda = 0.03$ is smaller. For the convenience of comparison the vorticity
plot of this study is rotated upside down.
The remainder of the leading vortex ring inside the droplet is visible in
all three examples. The features of the counter rotating vortex induced by
the earlier impinging wake can also be found in all vorticity plots shown. The
drainage of the film between the droplet and the interface is slow due to the
considerably lower viscosity ratio $\lambda = 0.03$ (i.e. larger ambient viscosity) in
Fig. \ref{pic15} a). A layer of ambient liquid is still draining from the liquid bridge
while the first minimum of the aspect ratio is reached. The remaing parts of
the leading vortex inside the drop are decelerated by this viscous film and
result in an area of counter rotating vorticity at the surface of the droplet.
Due to a limited field of view not the full area of interest upstream of the
droplet position d3 can be seen.

\subsection{Comparison of simulation and experiment}

In this section, results of the experimental investigation are compared to results
of the numerical study by Coyajee et al. \cite{coyajee2006direct}. In the numerical study, drop
impact on an interface is simulated using the MCLS method \cite{van2005mass}, a Finite difference/
Front-capturing method. To compute fluid motion, the Navier-
Stokes equations are solved on a fixed Cartesian grid, taking into account
different viscosity and density in each phase and surface tension forces on
the interface between the phases. The interface is tracked with a combined
Level-Set/Volume-of-Fluid method. For further details on the computational
method, we refer to \cite{vdpijlthesis}, \cite{van2005mass} and \cite{coyajee2006direct}.

\subsubsection{LIF results}

\begin{figure}
\includegraphics[scale=0.3]{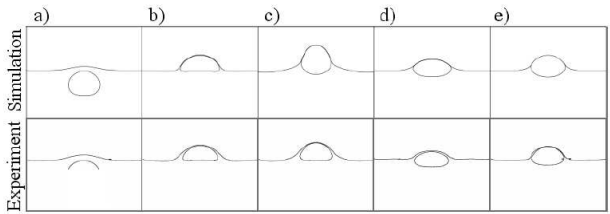}
\caption{Qualitative comparison of the computed (top) and the LIF measurement
of the Droplet shape (bottom).}
\label{pic16}
\end{figure}

In Fig. \ref{pic16}, contour lines of the droplet and the top layer interface are compared
at different instances during the impact event. The upper row shows
the numerical results, while the bottom row represents processed images of
the experimental LIF data. Note that the experimental data are obtained
within a limited field of view and therefore only part of the droplet is visible
in the first frame. Fig. \ref{pic16} reveals qualitative agreement of the droplet and
interface deformation during drop impact. However, in the simulation the
droplet is found to penetrate further into the top layer (image c). In addition,
the simulation shows immediate drainage of the bridging gap between
the droplet and the interface of the top layer. Due to a limited number of
computational cells available this thin film is not resolved in the simulation.

For quantitative comparison of experiment and simulation three parameters
are chosen:
\begin{itemize} 
\item vertical position of the droplet center of mass,
\item vertical velocity of the droplet center of mass,
\item droplet aspect ratio (ratio of the horizontal and vertical droplet diameter).
\end{itemize}

\begin{figure}
\includegraphics[scale=0.3]{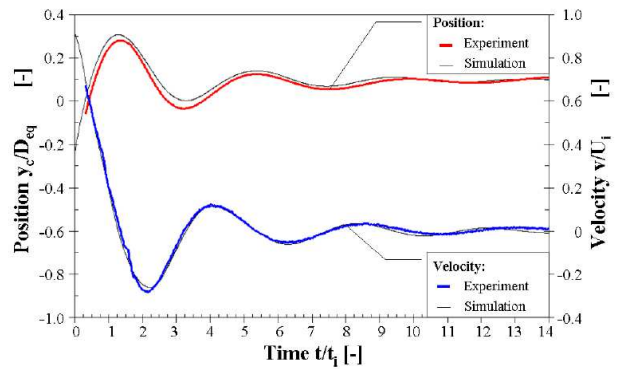}
\caption{Center point position and velocity of the simulation (thin line) is compared
to the experimental results (thick line).}
\label{pic17}
\end{figure}

\begin{figure}
\includegraphics[scale=0.4]{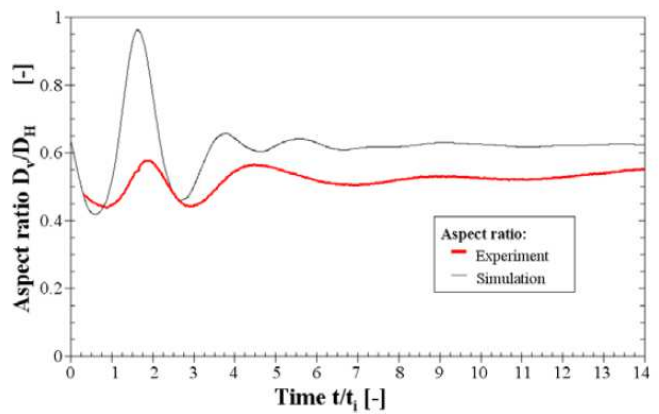}
\caption{Comparison of the aspect ratio of the droplet during the impact as obtained
from the experimental and numerical results.}
\label{pic18}
\end{figure}

Time evolution of the droplet position and velocity is presented in Fig. \ref{pic17}
for both the experiment and simulation. The left vertical axis of Fig. \ref{pic17} shows
the droplet position, while the right vertical axis shows the droplet velocity.
The simulation of the droplet center point matches the measurement data
closely.
In Fig. \ref{pic18}, time evolution of the droplet aspect ratio is displayed. Comparison
of simulated and experimental results shows good qualitative agreement,
but the magnitude of the droplet deformation is too large in the numerical
simulation.
The difference between the numerical and experimental results may be
explained from the importance of the thin film, which is not resolved in the
numerical simulation. In the current drop impact experiment the viscosity of
the continuous phase is much larger than the viscosity of the droplet phase.
As a result, a considerable film of the continuous phase persists in the gap

between the droplet and the top layer interface during the impact event. In
the study of Mohamed-Kassim \& Longmire, the viscosity ratio is much closer
to unity (Tab. 1) and the thin film between the droplet and the interface
is much smaller (Fig. 2,3 and 6 of \cite{mohamed2003drop}). Preliminary simulations using the
experimental parameters of Mohamed-Kassim \& Longmire show much better
quantitative agreement between numerical and experimental results for the
droplet deformation.

\subsubsection{PIV results}

\begin{figure}
\includegraphics[scale=0.4]{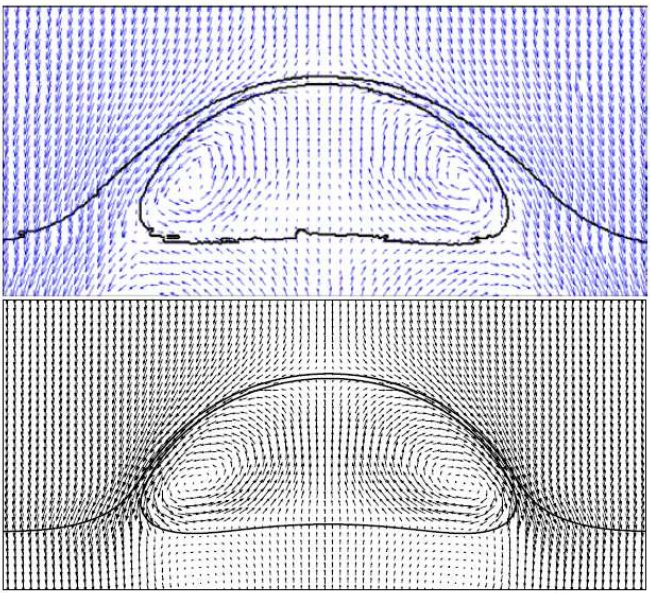}
\caption{Interfaces and velocity vectors at maximum $D_w/D_h$ of the experiment
(top) and the simulation (bottom).}
\label{pic19}
\end{figure}

In Fig. \ref{pic19}, flow fields of experiment and simulation are compared by displaying
velocity vectors over a cross section of the domain. The time instant
of each image is $t/t_i = 0.9$, where $D_w/D_h$ has a maximum. Qualitative
agreement of the flow patterns is found, e.g. the similarly situated center
of rotation inside the droplet, the impinging wake and the jet caused by
the draining film. The thickness of the film in the simulation is smaller and
almost constant along the interface of the droplet.

\begin{figure}
\includegraphics[scale=0.4]{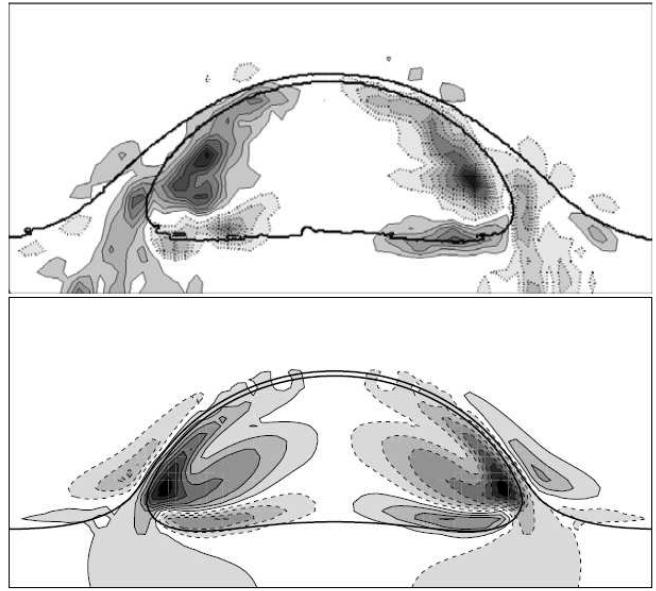}
\caption{Interfaces and vorticity contour levels at maximum $D_w/D_h$ of the experiment
(top) and the simulation (bottom).}
\label{pic20}
\end{figure}

The vorticity plots shown in Fig. \ref{pic20} are derived from the velocity fields
presented above (Fig. \ref{pic19}). The upper image represents the experimental data,
the lower shows the result of the simulation. Most flow features shown by the
experiment can also be found in the results of the DNS. The resemblance of
the vortex ring inside the simulated droplet and the measurement is good and
the vorticity due to the impinging wake matches. Also the shear field of the
drained fluid from the film is present in both the experiment and simulation.

\section{Conclusion}

This paper describes an approach to optimize the evaluation quality of highspeed
PIV. Spatial and temporal velocity gradients in the flow are the cause

of variations in the quality of the PIV result. Large time separation result in
decorrelation with an increase of spurious vectors as a result. A small time
separation provides a very high reliability (very few spurious vectors) but are
not meaningful because the measured particle image displacement is found to
be in the range of measurement error (∼ 0.1 pixel). Taking into account the
observed flow conditions, an adjustable time separation between two correlated
images is proposed. In order to find an optimal time separation within
a high-speed sequence the fraction of spurious vectors is used to monitor
the data quality. Based on an analysis of the flow properties with standard
evaluation methods the RMS of the particle image displacement of the vector
field is found to be an indicator for spatial gradients in the considered
flow. Keeping the RMS constant over time, we achieved a constant fraction
of spurious data and thus identified the largest possible time separation. This
maximized the VDR throughout the measurement sequence. Further work
has to be done in order to explore the possibility to use an adjustable separation
time not only related with entire images but also for each interrogation
window within an image frame. In this manner it will be possible to calculate
vector fields with an optimal particle image displacement throughout
the whole vector field series and within each field.
The comparison of the experimental data presented in this study with numerical
calculations presented by \cite{coyajee2006direct} reaches at this stage a basic level. The
parameters extracted from the experiment can only provide general statements
about the capability of the simulation to predict the flow pattern. The
qualitative and quantitative parameters chosen show a reasonable capture of
droplet shape and good results for the simulation of the center point motion
and velocity. The difference in aspect ratio may be partially explained with
the fast drainage of the thin film of viscous fluid in the simulation. However,
judging the capability of the numerical simulation may be topic of future
detailed investigation of the now improved PIV data set.


% \bibliography{lit}
% \bibliographystyle{plain}
\bibliography{lituli2}

\end{document}